\documentclass[journal]{IEEEtran}
\usepackage{graphicx,epsfig,amsmath,amssymb,bm}
\usepackage{amssymb,balance}
\usepackage{url}
\usepackage{hyperref}
\usepackage{breakurl} 
\usepackage{amsmath}
\usepackage{amsthm} 
\usepackage{amssymb}
\usepackage{algorithmicx}
\usepackage{algpseudocode}
\usepackage{cite}
\usepackage{url}
\usepackage{bbm}
\usepackage{subcaption}
\usepackage{tabularx}
\usepackage{multirow}
\usepackage{verbatim}
\usepackage[ruled,linesnumbered]{algorithm2e}

\usepackage{color}
\usepackage{booktabs}
\usepackage{tikz}
\usepackage[acronym]{glossaries}
\usepackage[capitalise,noabbrev]{cleveref}
\usepackage{enumitem,kantlipsum}
\usepackage{mathabx}
\usepackage{xurl}
\usepackage{caption}
\usepackage{makecell}

\begin{document}

\title{Survey of Load-Altering Attacks Against Power Grids: Attack Impact, Detection and Mitigation}
\author{Sajjad Maleki\IEEEauthorrefmark{1},~\IEEEmembership{Student Member, IEEE}, Shijie Pan\IEEEauthorrefmark{1},~\IEEEmembership{Student Member, IEEE},\\ Subhash~Lakshminarayana~\IEEEmembership{Senior Member, IEEE}, Charalambos Konstantinou~\IEEEmembership{Senior Member, IEEE} \vspace{-0.2 in}

\thanks{{S. Maleki is with the School of Engineering, University of Warwick, CV47AL, UK and ETIS UMR 8051, CY Cergy Paris Universit\'e, ENSEA, CNRS, F-95000, Cergy, France.}}
\thanks{{S. Lakshminarayana is with the School of Engineering, University of Warwick, Coventry, UK, CV47AL.}}
\thanks{{S. Pan and C. Konstantinou are with the Computer, Electrical and
Mathematical Sciences and Engineering Division, King Abdullah University
of Science and Technology, Thuwal 23955, Saudi Arabia.}}

\thanks{{*Equal contribution. Author ordering determined by coin flip.}}
}

\maketitle

\begin{abstract}

The growing penetration of IoT devices in power grids despite its benefits, raises cybersecurity concerns. In particular, load-altering attacks (LAAs) targeting high-wattage IoT-controllable load devices pose serious risks to grid stability and disrupt electricity markets. This paper provides a comprehensive review of LAAs, highlighting the threat model, analyzing their impact on transmission and distribution networks, and the electricity market dynamics. We also review the detection and localization schemes for LAAs that employ either model-based or data-driven approaches, with some hybrid methods combining the strengths of both. Additionally, mitigation techniques are examined, focusing on both preventive measures, designed to thwart attack execution, and reactive methods, which aim to optimize responses to ongoing attacks. We look into the application of each study and highlight potential streams for future research.

\end{abstract}
\vspace{-1mm}

 {\IEEEkeywords Load-altering attacks, impact analysis, detection, localization, attack mitigation, survey.}

\IEEEpeerreviewmaketitle
\vspace{-2mm}
\section{Introduction} \label{Introduction}
\vspace{-2mm}
Power grids are witnessing a rapid proliferation of smart energy appliances, such as smart electric vehicle charging stations (EVCS) and smart heat pumps. The connectivity features of these devices can be used to schedule their usage, hence offering demand flexibility (DF). For instance, in the United Kingdom (UK), the government is backing the expansion of public EV charging infrastructure to reach $300$k public chargers by 2030 with an investment totaling \pounds$1.6$b \cite{ukgov2024}, while the capital of Saudi Arabia, Riyadh, targets towards rising the EV fleet to over $1$m by 2030 \cite{riyadhEV}.

However, these IoT-enabled appliances can become a serious source of security vulnerability to target the power grid. Several security issues, such as insecure authentication, lack of firmware security upgrades, etc., have already been discovered in these devices (see Section \ref{threat_modeling} for a detailed discussion). {Overall,} they can be particularly attractive targets for malicious actors due to the following factors: \textit{(i)} the security standards for these devices are still in their infancy, and there is a lack of a unified approach for securing them; \textit{(ii)} they are typically manufactured and installed by third parties, who may typically ignore security features/recommendations due to cost considerations; \textit{(iii)} the power grid operator does not have direct control over these devices; \textit{(iv)} they are operated by end-user customers who may lack security awareness. 

One major threat emerging from these vulnerabilities is the concept of load-altering attacks (LAAs) \cite{mohsenian2011distributed}. LAAs exploit insecure, high-wattage IoT-enabled devices, such as EVCS, smart thermostats, and industrial loads, to manipulate power consumption on a large scale. By coordinating synchronized load changes, attackers can disrupt the power grid’s frequency stability, induce voltage fluctuations, and, in extreme cases, cause cascading failures or blackouts. Unlike traditional cyberattacks targeting SCADA systems or distributed energy resources (DERs), LAAs manipulate end-user loads directly, which makes them harder to detect and mitigate.

The threats go beyond mere theoretical constructs. Real-world cyberattacks involving smart appliances have already been witnessed. For instance, an EVCS was hacked to display inappropriate images \cite{bbc2022}. There have also been reports of compromising and disabling of EVCS near the Moscow region and hackers using their displays to show political messages \cite{darkreading2023}. 
In the UK, EV chargers from a specific vendor were pulled out of the market due to inadequate cybersecurity features in its design, with the concern that the chargers can be {exploited} as a means to target the power grid \cite{telegraph2024}. In December 2023, a US court-authorized operation thwarted a botnet attack targeting several small office/home office routers against a ``campaign to target critical infrastructure organizations in the US'' \cite{justice2024}. 

\begin{figure}
    \centering    \includegraphics[width=0.8\linewidth]{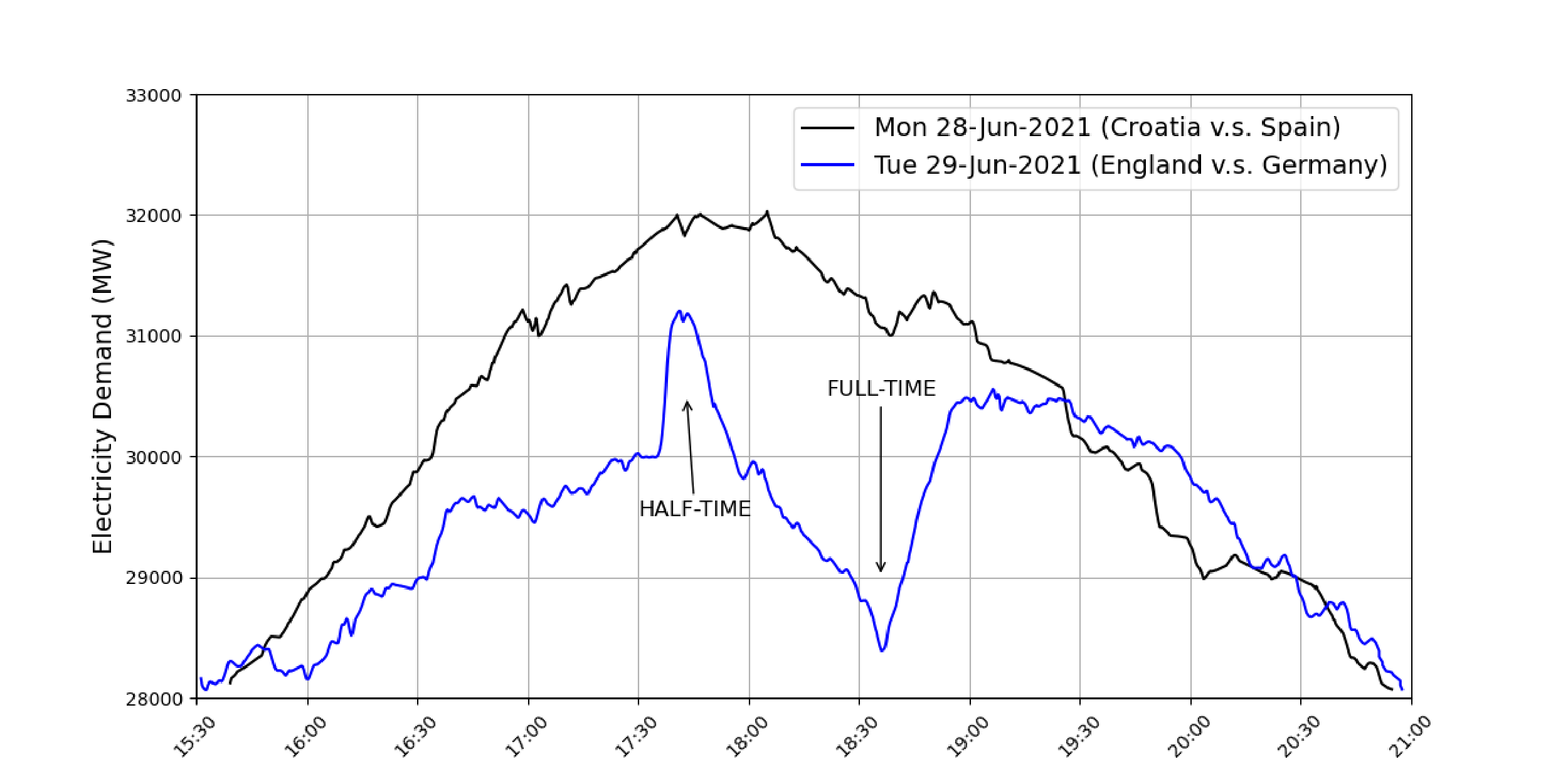}
    \vspace{-2mm}
    \caption{TV Pickup effect observed in UK power grid  \cite{TVPickupNGESO}.}
    \vspace{-4mm}
    \label{fig:TVPickup}
\end{figure}

\begin{figure}[t]
        \centering
    \includegraphics[width=0.49\textwidth]{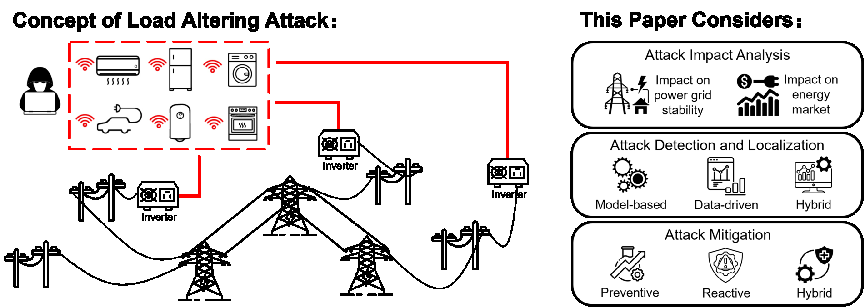}
    \vspace{-5mm}
    \caption{The concept of load-altering attacks (LAAs) and the discussed topics within this survey paper.}
    \vspace{-5mm}
    \label{Structure}
\end{figure}

The focus of this work is large-scale compromises, such as a botnet attack targeting a large number of IoT-load devices {i.e., LAAs}. A popular instance is the Mirai botnet attack which at its peak targeted $600k$ IoT devices. A real-world illustration of synchronously turning ON/OFF a large number of consumer electrical appliances can be observed during the so-called ``TV Pickup effect'' experienced by the UK's power grid \cite{TVPickupNGESO}. It refers to large-scale load surges or drops in the load during popular football matches. During half-time and full-time breaks, it has been observed that customers synchronously turn on electric kettles or open their fridge doors, the aggregate effect of which resulted in several GW of load change, as illustrated in Fig.~\ref{fig:TVPickup}. If such events occur due to a cyber attack, they will remain unanticipated, resulting in grid stability risks. Table \ref{CVE} represents a list of known common vulnerabilities and exposures (CVE) which provide entry points for adversaries to execute such attacks.

\begin{table*}
    
    \centering

    \caption{A list of known Common Vulnerabilities and Exposures (CVEs) related to potential LAA threats.}
    \vspace{-1mm}
    \begin{tabular}{|p{2cm}|p{2.5cm}|p{11cm}|} \hline

    \multirow{3}{2cm}{High Wattage Device\centering} & Common Vulnerabilities and Exposures\centering & \multicolumn{1}{c|}{\multirow{3}{*}{Description}} \\    
    \hline 
    \centering \multirow{2}{*}{EV charger} & \centering CVE-2022-0878 & ``Brokenwire'' attack: disrupt EVCS by interfering communication between EVs and chargers. \\ \cline{2-3} & \centering CVE-2021-22774& Requiring the possibility of accessing knowledge about the charging station user account.\\\hline
     
    \centering \multirow{4}{*}{Air conditioner} & \multirow{1}{2.5cm}{\centering CVE-2021-20595} & Allows unauthenticated attackers to disclose the data or cause a DoS via sending crafted packets.\\ \cline{2-3} & \centering CVE-2022-24296 & Provides the possibility of encrypted messages revealing to unauthorized entities.\\ \cline{2-3}& \centering CVE-2022-33322 & A distant attacker can launch a script in the users' interface to expose secure data. \\ \cline{2-3} & \centering CVE-2021-20593& An exposure which enables the attacker to uncover and manipulate device configurations.\\\hline
    \centering Heat pump & \centering CVE-2024-22894 & Provide the possibility of launching arbitrary codes for the attacker.\\ \hline
    \centering \multirow{3}{*}{Inverter} & \multirow{2}{2.5cm}{\centering CVE-2023-0808} & An issue in the access point setting handler component, where manipulation of input could result in the use of a hard-coded password, potentially enabling attacks on the physical device. \\\cline{2-3} 
    & \multirow{1}{2.5cm}{\centering CVE-2023-39169} & Impacted devices utilize pubicly-available default credentials granting administrative privileges. \\ \hline
    \centering \multirow{1}{*}{LED TV} & \multirow{1}{2.5cm}{\centering CVE-2023-4614} & Remote attackers can execute a code on impacted installations without authentication.\\

    \hline
    \end{tabular}
        \label{CVE}
\vspace{-6mm}
\end{table*}

\vspace{-5mm}
\subsection{Difference From Existing Works}
\vspace{-1mm}
The topic of power grid security has attracted significant interest from the research community in the last decade and several survey papers have been written on this topic \cite{MashimeNOW2023, zhang2021smart, ghiasi2023comprehensive, sayghe2020survey, abdi2024role}. A vast majority of the works focus on cyberattacks against the power grid's supervisory control and data acquisition (SCADA) system. 
Notable examples include false data injection attacks (FDI), denial of service (DoS) or time delay attacks targeting the data communicated between power grid sensors and the control centers, physical attacks that alter the power grid topology (e.g., remotely opening transmission line circuit breakers), coordinated data and physical attacks, time synchronization attacks against phasor measurement units (PMUs), etc. The power grid industry and policymakers have taken note of these threats and developed security solutions to counter them and already paved their way to form regulations/standards, such as the IEC 62351 standard which provides security recommendations for the IEC 61850 protocol \cite{hussain2019review}. Furthermore, in the research literature, several cyber-physical defense solutions are proposed in existing papers, such as strategic protection of sensors to prevent undetectable attacks \cite{kim2011strategic}, machine learning (ML)-based approaches to detect cyberattacks \cite{sayghe2020survey}, and moving target defense to invalidate attacker's knowledge of the target system and thus making the attacks detectable \cite{lakshminarayana2024survey}.

While the papers discussed before address the threats against the bulk power grid and centralized generations and their control loops \cite{sridhar2014model, tan2016optimal, ameli2018attack}. However, with the high integration of the DER in the grids and the vulnerability of these resources against cyberattacks, recent works have turned their attention to studying attacks on distributed and decentralized generations (similar to LAAs which are attacks on edge devices which are scattered all over the grid). In \cite{zografopoulos2023distributed, liu2024enhancing, ye2021review, li2022cybersecurity}, the main focus is on cyberattacks targeting DERs and their components. Reference \cite{zografopoulos2023distributed} surveys the security of DERs in both cyber and physical layers and covers the attacks targeting DERs and their communication means while \cite{liu2024enhancing} discusses vulnerabilities and cyber-resiliency of DER-based grids. Furthermore, \cite{ye2021review} makes a comprehensive overview on attacks on photovoltaic systems and their mitigation methods. The work in \cite{li2022cybersecurity} provides an extensive survey of existing studies on the cybersecurity of smart inverters of DERs. Overall, these works primarily address threats that affect DER operations, such as manipulating DER protocols or exploiting inverter-level weaknesses to disrupt generation or storage capabilities. 
Unlike DER-based attacks that target the generation and storage aspects of power systems, LAAs manipulate the load directly through IoT-enabled smart appliances, disrupting the balance between power supply and demand. 

While DER-based attacks compromise the integrity and control of distributed resources, LAAs exploit vulnerabilities in consumer-side appliances, leading to fluctuations in grid stability and potentially causing frequency deviations and cascading failures. This distinction underscores the need for dedicated research on LAAs, as they pose unique threats by targeting end-user devices. To the best of our knowledge, this work is the first comprehensive review of LAAs against power grids.



A few recent survey papers have investigated cyber threats from end-user devices. For instance, \cite{asghar2017smart} surveys the privacy of smart meters. However, according to the lack of a comprehensive review paper specifically focusing on LAAs, this survey paper is dedicated to LAAs, targeting the end-user smart energy appliances that are installed at the customer end as depicted in Fig. \ref{Structure}.

To summarize, the contributions of this paper are as follows: 
\begin{itemize}
    \item We provide a comprehensive survey of risks due to LAAs, highlighting the threat modeling and analyzing the impact of LAAs on power grid operations as well as the impact on energy markets. 
    \item We present a review of existing mechanisms to detect and localize LAAs, including model-based, data-driven, as well as hybrid methods.
    \item We provide an extensive overview of works on how to  mitigate and recover from LAAs, including preventive and reactive mitigation schemes.
    \item Finally, we conclude the survey by presenting an overview of the future research directions in this field. 
\end{itemize}

The rest of the paper is organized as follows: Section \ref{threat_modeling} discusses the threat modeling, Section \ref{Impact} surveys the impact of the attack followed by studying attack detection and localization frameworks in Section \ref{detection}. Section \ref{Mitigation} presents the mitigation methods for LAAs and eventually, Section \ref{conclusion_future_research} remarks on the future research and concludes the paper.

\vspace{-3mm}
\section{Threat Modeling} \label{threat_modeling}
\vspace{-1.5 mm}
Direct system-end attacks targeting the assets in the SCADA system are the most prevalent form of cyber threats (such as those noted in Section~I). Conversely, the deployment of DERs at the consumer level along with the large-scale deployment of high-wattage IoT devices introduces new vulnerabilities, enabling adversaries to orchestrate and execute remote cyberattacks. For instance, DERs leverage IoT technology to synchronize their operational controls with power systems. Notably, devices such as smart inverters are particularly prone to cyber-threats. When compromised, these devices can interfere with the functioning of the grid, disrupt the delivery of electricity, and potentially destabilize the entire grid \cite{9351954, KURUVILA2021107150, LakshCOVID2022}. 

Unlike traditional power generation facilities, DERs are managed by end-users and aggregators, limiting utility companies' control over electricity production. The IoT interfaces utilized by these DERs and other high-wattage consumer equipment are particularly vulnerable to cyberattacks due to their complex supply chains and limited resources. Such vulnerabilities make them prime targets for various cyber threats. For instance, insecure remote login credentials can be leveraged to spread malware like the Mirai botnet \cite{7971869}. Additionally, the process of updating firmware presents another risk for introducing malicious software \cite{Cui2013WhenFM}. Furthermore, Ghost domain name system (DNS) attacks pose a significant threat as they can redirect IoT devices to harmful DNS servers \cite{Jiang2012GhostDN}. The security of electricity consumers and DERs is also compromised by the frequent use of default or weak passwords, leading to an increased risk of cyberattacks.

This paper examines the threat landscape of indirect consumer-level cyberattacks on the power system via load alternation. Such attacks are possible because DER-based inverters and IoT-based high-wattage devices have communication and control interfaces which lack the advanced cyber-defense and can be exploited with the malicious intent to destabilize the grid. Many attack vectors use a multi-layer threat model by leveraging the complexity and interconnections within the power grid. As shown in previous works \cite{CardenasHahnIoT2020, konstantinou2019hardware, konstantinou2015impact}, attacks can target one or multiple layers of the IoT-connected appliances and DERs. Adversaries may opt to infiltrate any part of the supply chain to inject stealthy trojans (i.e., malicious and deliberate modifications, consisting of a multi-layer model of attack vectors aiming to compromise, control, modify, or disable the system or leak sensitive information) into the IoT firmware/software. The attacks can be either physical (i.e., access required to extract device code) or network-based (e.g., remote over-the-air manipulation). LAAs emphasize the cross-layer propagation of such attacks which can cause significant load alteration. 
 
\vspace{-4mm}
\subsection{Attacker Model}
\vspace{-1mm}
Using IoT-related vulnerabilities, an attacker would aim to modify control and operation actions of IoT-load devices under normal conditions to actions that would harm DERs and/or distribution and transmission systems. Following this modification, the production and consumption of each DER/IoT-load are changed and, as a result, the power demand at those nodes changes. Using this mechanism, the attacker will aim to coordinate actions of compromised devices in the system to maximize the potential damage. The attack model of LAAs assumes that the attacker possesses some, but not necessarily exact, knowledge of the system design (to drive the system conditions into activating specific protection and control actions) e.g., topology, parameters of lines, buses, generators, etc, which can be obtained from publicly available datasets (e.g., \cite{conedison_2018, 7039273, acharya2020public, keliris2019open}), while further details can be inferred with a high accuracy  \cite{7810304, 7914787}.

First, the knowledge of network topology, line parameters, bus data, generator configurations, and EVCSs can be deduced from publicly available datasets \cite{acharya2020public}.
Examples include collecting locations and power ratings of EVCSs, real-time and historical usage data from EV smartphone applications, locations of generators, transmission lines, transmission substations, and their capacity from public repositories such as the U.S. Energy Information Administration, and real-time generation and demands from NYISO datasets \cite{acharya2020public}. 
However, some specifications like the length of lines are estimated, and may not reflect the true values.  Using such gathered intelligence, 
\cite{soleymani2024data} design an (EV-oriented) LAA which constructs the attack vector based on the grid's frequency only and still causes a destructive impact.

Secondly, topology identification methods allow attackers to deduce network structures based on observable measurements in both transmission and distribution networks \cite{francis2022enhancing, grotas2019power}.
Specifically, in distribution networks, an attacker can deduce the network topology by accessing the nodal voltage measurements and applying signal processing/machine learning (ML) techniques such as deep neural network \cite{li2023distribution}, linear discriminant analysis and regularized diagonal quadratic discriminant analysis \cite{francis2022enhancing}, generative adversarial networks \cite{wu2022gridtopo}, and recursive grouping algorithm s\cite{park2018exact}. As distribution grids are sparsely monitored, attackers can also use measurements from advanced metering infrastructure (AMI) to derive the network topology  \cite{liang2021power},  \cite{pappu2017identifying}. 
Similar signal processing/ML-based approaches can also be used to deduce the topology of transmission networks \cite{grotas2019power}, \cite{li2013blind}.

Existing literature often formalizes the attack model of LAAs by co-optimizing it with the distribution system operations as given in and adopting a common cybersecurity practice, which is to conservatively assume a strong (omniscient) attacker and a weak (naive) defender (e.g., DSO) \cite{chen2020load, li2023dynamic, youssef2023adversarial}. Accordingly, the omniscient attacker has full knowledge of the system, while the naive defender has no detection or defense mechanisms to respond. These assumptions lead to assessing the worst-case impact of the attack. 
The omniscient attacker can pursue instant and lasting effects on the power system and DERs, given oracular knowledge of the system conditions. The attacker objective can encompass: \textit{(i)} reducing efficiency of DSO operations by reducing the energy and cost performance of the system, \textit{(ii)} equipment damages and failures by incurring additional wear-and-tear due to violating operation limits on equipment and intelligently triggering relay protection to cause load shedding and blackouts, \textit{(iii)} stressing the system during extreme events (e.g., natural disasters or peak conditions) to increase the likelihood of equipment failures. 
    
The model considered in the works above analyzes the worst-case impact of cyberattacks, i.e., under the assumption of the omniscient attacker and naive defender. In practice, it is unrealistic that an attacker can monitor every aspect of DER or power system operation. Therefore, it is realistic to account for the attacker's limited knowledge of actions of uncompromised DERs and IoT-loads, and the DSO, which prevents the attacker from precisely estimating attack impacts on the grid and from incurring the theoretical maximum damage. The attacker's limited knowledge (constraints) can be considered in various manners, including: 

\textit{(i) Statistical modeling of uncompromised DERs/IoTs}: using open-source data sets,  e.g., Pecan Street \cite{pecan_project} or Tracebase data repositories \cite{tracebase_project}, the attacker can construct a statistical model to describe the behavior of uncompromised devices. These repositories have sufficient data to estimate the mean and standard deviation, which can be used to parameterize the uncompromised devices via the Gaussian distribution, or more complex statistical models (e.g., to account for temporal and spatial correlations \cite{9642055}). Under such statistical models, the threat model of LAAs can be reformulated into different instances of computationally tractable robust optimization models, which can be used to  formalize and compare different attack objectives under various statistical assumptions, i.e., quantify realistic rather than the worst-case attack effects. 

\textit{(ii) Adversarial Modeling of the Response}: Building on statistical models, one can consider an adversarial behavior against the DSO, which renders a game-theoretic formulation for DSO-attacker interactions.  This adversarial modeling can provide a realistic rather than naive grid response and  can be implemented in a static or dynamic manner, where the attacker fully or partially observes the system response and modifies the attack accordingly \cite{HOTA2016135}.  This can 
determine the existence, uniqueness, and properties of the equilibrium for single/multi-stage (sequential)  games using computationally scalable algorithmic game theory  \cite{nisan2007algorithmic, maleki2024distribution}.  

\textit{(iii) Learning for Partial System Observability}: The statistical and adversarial models described above, can be supported by considering the limited observability of the system by the attacker  (e.g., it can only partially observe  power flows, voltages, and other system parameters). In this case, the attacker's problem can be formulated via reinforcement learning (RL), where the observed nodal parameters are the environmental states and the attack damage is the reward. Using such  RL frameworks can lead to continuous learning, instead of the omniscient attacker, in a multi-stage adversarial game. 

\vspace{-2mm}
\subsection{Attack Types}

\begin{table*}
    \centering
    \vspace{-2mm}
    \caption{Classification of LAAs.}
    \vspace{-2mm}
    \begin{tabular}{|c|c|c|c|} \hline
    Criterion&Category&Description&Works\\   
    \hline
    \multirow{3}{*}{Attack Scope} & Localized &  \makecell{Attacks are designed to specifically target and disrupt a\\ particular region, bus, or feeder within the power grid.} & \makecell{\cite{acharya2020public,jahangir2024charge,  HammadSwitch2018, amini2016dynamic,  HuangUSENIX2019, maleki2024impact,  LaksRareEvent2022}}  \\
    \cline{2-4}
    &Multi-area&  \makecell{Attacks are designed to compromise and manipulate \\loads across multiple locations.} & \makecell{\cite{ LakshIoT2021, Soltan2018, ospina2020feasibility,   Goodridgecascade2023, dabrowski2017grid, goodridge2024uncovering,chen2023enhancing, WeiEVSEBotnet2023, khan2019impact, LakshCOVID2022, mohsenian2011distributed, ospina2023feasibility, barreto2014cps, shekari2021mamiot, LiuSoftOp2022}} \\
    \hline
    \multirow{3}{*}{Feedback Type} & Static LAAs &  \makecell{Attackers alter loads once or at fixed intervals without \\ real-time monitoring of grid response.}& \makecell{\cite{acharya2020public,  LiuSoftOp2022, maleki2024impact, khan2019impact, LakshCOVID2022, mohsenian2011distributed, LakshIoT2021, Soltan2018, HuangUSENIX2019, ospina2020feasibility,  LaksRareEvent2022,   Goodridgecascade2023, dabrowski2017grid, goodridge2024uncovering}} \\
    \cline{2-4}
    &  Dynamic LAAs &  \makecell{Attackers continuously adjust loads in response to  \\system conditions that can be measured}& \makecell{\cite{amini2016dynamic,  dabrowski2017grid, goodridge2024uncovering, chen2023enhancing, WeiEVSEBotnet2023, Goodridgecascade2023}}\\
    \hline
    \multirow{3}{*}{Control Mechanism} & Informed-Knowledge &  \makecell{Attackers possess detailed grid parameters, including \\load profiles, grid topologies, or system responses.}  &  \makecell{\cite{acharya2020public,  LiuSoftOp2022, maleki2024impact, khan2019impact, LakshCOVID2022, mohsenian2011distributed, LakshIoT2021, Soltan2018, HuangUSENIX2019, ospina2020feasibility,  LaksRareEvent2022,   Goodridgecascade2023, dabrowski2017grid, goodridge2024uncovering,chen2023enhancing, WeiEVSEBotnet2023,HammadSwitch2018, amini2016dynamic,ospina2023feasibility, barreto2014cps, shekari2021mamiot, jahangir2024charge}} \\
    \cline{2-4}
    & Limited-Knowledge & \makecell{Attackers have partial or estimated knowledge of the \\ power system, (e.g., frequency measurements).} & \makecell{\cite{soleymani2024data}} \\
    \hline
    \multirow{5}{*}{Targeted System} & Transmission Grid &  \makecell{Attacks aim to destabilize grid frequency or induce \\ cascading failures in the transmission grid.} & \makecell{\cite{LakshCOVID2022 , mohsenian2011distributed, LakshIoT2021, amini2016dynamic, chen2023enhancing, WeiEVSEBotnet2023, Soltan2018, HuangUSENIX2019, ospina2020feasibility,dabrowski2017grid,LaksRareEvent2022,goodridge2024uncovering,Goodridgecascade2023,acharya2020public, HammadSwitch2018, AlanaziLoadOsc2023}} \\   
    \cline{2-4}
    &Distribution Grid &  \makecell{Attacks cause localized voltage instability or disrupt \\ demand response programs in the distribution grid.} & \makecell{\cite{LiuSoftOp2022, maleki2024impact, khan2019impact}}\\
    \cline{2-4}
    &Energy Markets &  \makecell{Attacks manipulate the energy market by artificially \\inflating or deflating demand for financial gain.} & \makecell{\cite{ospina2023feasibility, barreto2014cps, shekari2021mamiot, jahangir2024charge}}\\
    \hline
    \end{tabular}
    \vspace{-4mm}
    \label{LAA_classification}
\end{table*}

Table \ref{LAA_classification} categorizes LAAs based on four key criteria: feedback type, attack scope, control mechanism, and targeted system. This classification provides insight into the diverse strategies used by adversaries and their impact on grid stability and energy markets. The feedback type distinguishes between static LAAs, e.g., \cite{LakshIoT2021, ospina2020feasibility, Goodridgecascade2023, LaksRareEvent2022, Soltan2018}, where load modifications occur at fixed intervals without real-time monitoring, and dynamic LAAs, e.g., \cite{chen2023enhancing, WeiEVSEBotnet2023, goodridge2024uncovering}, where attackers adjust their actions based on system response, making detection more challenging. The attack scope differentiates between localized attacks, which affect specific buses or feeders, and multi-area attacks, which manipulate loads across multiple regions, potentially causing widespread grid instability. The control mechanism is classified into informed-knowledge attacks, where adversaries have detailed grid parameters, and limited-knowledge attacks, where attackers rely on partial system insights, such as frequency measurements \cite{soleymani2024data}. Finally, the targeted system classification highlights the different objectives of LAAs, including destabilizing the transmission grid by inducing frequency deviations and cascading failures, e.g., \cite{chen2023enhancing, WeiEVSEBotnet2023, goodridge2024uncovering}, disrupting the distribution grid through voltage instability and demand response manipulation \cite{LiuSoftOp2022, maleki2024impact, khan2019impact}, and exploiting energy markets by artificially inflating or deflating demand for financial gain \cite{ospina2023feasibility, barreto2014cps, shekari2021mamiot, jahangir2024charge}. This classification framework underscores the multifaceted nature of LAAs, providing a deeper understanding of their potential impact on grid operations and market dynamics.

In conclusion, vulnerabilities of IoT-enabled load devices are exacerbated due to complex supply chains and limited security in consumer devices, allowing attackers to launch sophisticated and multi-layered LAAs. The attacker can leverage partial system knowledge to modify control actions of IoT and DER devices, potentially destabilizing the grid. Realistically, the attacker's knowledge is constrained, making statistical, adversarial, and RL models relevant for capturing the nuanced threat dynamics. These models highlight that, despite the limitations in system knowledge, attackers can still inflict significant harm by exploiting communication and control interfaces in IoT-connected devices, thereby requiring robust cybersecurity measures at both the consumer and grid levels.

\vspace{-3mm}
\section{Attack Impact Analysis} \label{Impact}
\vspace{-1mm}

\subsection{Impact on Power Grid Operation and Stability} \label{Impact_stability}

\vspace{-0.1 cm}
Understanding the impact of LAAs on grid operations and identifying the most disruptive attack points are crucial for risk analysis. Table \ref{LAA_impacts} summarizes the existing studies on this topic. This table highlights the models studied, and the attack impacts considered in each work.

\subsubsection{Mechanism of Grid Operations}
An unforeseen load surge or drop can cause frequency deviations. The primary frequency response, provided by the system's inertia and fast-acting technologies like battery storage, compensates for these deviations in the first few seconds of an incident. Secondary and tertiary frequency responses are additional components of frequency control schemes that operate more slowly and are supported by resources like automatic generation control (AGC), spinning reserves, mobile batteries, and off-grid DERs.
In distribution systems, such load changes can result in voltage constraint violations. To mitigate this, devices such as shunt capacitors, batteries, flexible AC transmission system (FACTS) devices like soft open points (SOPs), circuit breakers, and DERs can be deployed. During LAAs, the adversary forces sudden, unexpected load changes, potentially disrupting {either} or both of the discussed equilibria.

\subsubsection{LAA Impact} To the best of our knowledge, the concept of LAAs was first introduced in \cite{mohsenian2011distributed}. 
The impact of sudden and large-scale load manipulations, referred to as static LAAs, on transmission systems was first studied using simulation-based approaches. {Studies} \cite{Soltan2018} and \cite{dabrowski2017grid} demonstrated that static LAAs can disrupt the equilibrium between generation and demand, resulting in frequency safety violations. A sudden increase in demand forces the grid operator to rely on primary generator response, potentially leading to increased line power flows. Since the operator lacks control over power flow during the primary response stage, excessive power flows could cause line outages or even cascading failures. Additionally, LAAs can affect steady-state metrics such as operational costs, as operators must activate more expensive ancillary services to manage increased demand.

Not only can attacks on transmission grids cause disruption in them, but also the damage of a well-crafted LAA on the distribution system can be propagated into the transmission grid. According to \cite{dvorkin2017iot}, a naive LAA on the distribution level does not consider the settings of the circuit breaker connecting the distribution system to the transmission grid. As a result, these attacks may trigger the circuit breaker and stop the propagation of the attack impact to the transmission level. However, in an LAA with an insidious strategy, the attacker accounts for the circuit breakers' settings and confines the resulting additional active and reactive power to stop activating the circuit breaker. This results in the propagation of the attack impact in the connected upstream transmission grid. 

While early studies provided valuable insights, they overlooked the impact of inherent protection measures in power systems. To address this, \cite{HuangUSENIX2019} analyzed the effects of LAAs considering protections like relays, load shedding, and $N-1$ scheduling. This realistic assessment showed that despite these protections, sudden load increases could still cause outages by triggering under-frequency load shedding or forcing islanded operation. However, these measures can prevent large-scale cascading failures, highlighting trade-offs in system protection.

Recent studies have also explored how system conditions, particularly low-load and low-inertia circumstances, influence the impact of LAAs. For instance, \cite{ospina2020feasibility} analyzed load consumption patterns across seven U.S. cities during the COVID-19 pandemic and demonstrated that LAAs targeting regions with high load proportions have the most severe impact on frequency stability. Motivated by these findings, \cite{LakshCOVID2022} explored the heightened vulnerability of power grids under low inertia conditions, such as those seen during the pandemic, where errors in RES forecasting can exacerbate the effects of LAAs. Their results showed that under these conditions, attackers can destabilize the system more easily, as low inertia leads to larger frequency swings which further complicates grid management for operators.

While static LAAs are concerning, dynamic LAAs (DLAAs) represent a more sophisticated and potentially more impactful form of attack. In a DLAA, an attacker repeatedly alters loads in an oscillatory manner, guided by the feedback signals from the grid. One variant, as proposed by \cite{amini2016dynamic}, involves load changes proportional to the local frequency deviation from its nominal value, thus exacerbating the load-generation imbalance strategically. However, continuous load changes can be difficult to implement in practice. Alternately, \cite{HammadSwitch2018} and \cite{AlanaziLoadOsc2023} explored DLAAs where load devices are switched on and off in synchronization with inter-area oscillatory frequencies, effectively exciting oscillations and driving the system toward instability.

To further analyze both static and dynamic LAAs, \cite{LakshIoT2021} developed an analytical framework leveraging second-order dynamical system theory. Their study derived eigenvalue sensitivities of the grid's frequency control loop and proposed a method to identify the nodes from which the attacker could launch the most efficient DLAAs. However, many of these works focus on specific LAA scenarios, such as LAAs of defined magnitudes at specific nodes. To address the broader landscape of potential attacks, \cite{LaksRareEvent2022}, \cite{Goodridgecascade2023}, and \cite{goodridge2024uncovering} employed rare-event sampling techniques to evaluate the impact of LAAs across various combinations of nodes and load magnitudes. Their findings emphasize that DLAAs occurring at discrete intervals, in combination with existing protection measures, can still lead to cascading failures, rendering dynamic LAAs more disruptive than static ones.

Another emerging focus is on LAAs targeting EVCS. The work in \cite{acharya2020public} demonstrated that destabilizing EVCS-based LAAs can be executed using publicly available data, such as EVCS locations and grid demand datasets. It proposed a data-driven load manipulation strategy that shifts system eigenvalues to a vulnerable region defined by NERC standards, without requiring real-time grid monitoring, unlike previous works \cite{amini2016dynamic, murguia2018reachable, AminiIdentification2019, peng2019survey}. Sayed \textit{et al.} \cite{sayed2022electric} conducted a case study highlighting the significant impact of EVCS-based LAAs due to high reactive power demand compared to residential loads.  \cite{WeiEVSEBotnet2023} examined coordinated DLAAs using distributed EVCS, factoring in communication delays and countering transient energy-based methods. Their robust optimization strategy maximized attack impact by coordinating multiple EVCS nodes.  \cite{sarieddine2023investigating} explored vulnerabilities in mobile apps controlling EV charging, showing through real-time testbed simulations that inadequate vehicle authentication allows remote attackers to manipulate charging operations and launch attacks.

While most papers has focused on LAAs targeting transmission networks, recent work has begun to address their impact on distribution networks. The authors in \cite{LiuSoftOp2022} identified issues like low voltage and line overloads in distribution networks due to LAAs and suggested soft open points as a mitigation strategy (discussed in Section \ref{Mitigation}). Maleki \textit{et al.} \cite{maleki2024impact} introduced an analytical framework with closed-form expressions to quantify LAA impacts, showing that attacks at downstream nodes (e.g., leaf nodes) are most effective, and highlighted the role of voltage-dependent loads (using the ZIP model). \cite{khan2019impact} investigated the voltage profile effects of EVCS-based LAAs on distribution networks. Additionally, \cite{dvorkin2017iot} studied the joint impact on T\&D networks using a bi-level optimization model, linking attack impact to IoT penetration and strategies, and noted key metrics like energy not served and increased costs, emphasizing the need {to understand} T\&D interplay for effective LAA risk mitigation.

\begin{table*}
    \centering
    \vspace{-4mm}
    \caption{Overview of the existing works on the impact of LAA on power grid operation and stability.}
    \vspace{-2mm}
    \begin{tabular}{|>{\centering\arraybackslash}p{0.4cm}| c |>{\centering\arraybackslash}p{4.1cm}| >{\centering\arraybackslash}p{2cm} |>{\centering\arraybackslash}p{3.8cm} |>{\centering\arraybackslash}p{3.6cm}| } \hline
    &Ref & \centering Model studied & Study type &\multicolumn{1}{c|}{Attack type} & {Analyzed impact} \\ \hline
    \multirow{14}{*}{\rotatebox[origin = c]{90}{Transmission}} & \cite{LakshCOVID2022 , mohsenian2011distributed, LakshIoT2021} & {$2^{nd}$ order model- DC PF \centering} &  Analytical &  {Static}  &  {Frequency instability} \\ \cline{2-6}
     &\cite{amini2016dynamic, chen2023enhancing, WeiEVSEBotnet2023} & {$2^{nd}$ order model- DC PF \centering}& Analytical & {Dyn. - Cont. load change} & {Frequency instability} \\  \cline{2-6}
    &\multirow{2}{*}{\cite{Soltan2018}}  & \multirow{2}{4.5cm}{\centering MATPOWER and PowerWorld \centering}&\multirow{2}{*}{\centering Simulation-based} &\multirow{2}{*}{\makecell{Static}} &Freq. Instability, Line Failures, Inc. Op. Costs\\ \cline{2-6}
         &\multirow{1}{*}{\cite{HuangUSENIX2019}} &PowerWorld + $(N-1)$ constr. & {\centering Simulation-based}& {\centering Static} & Freq. instab., Line overloads\\ \cline{2-6}
     &\multirow{1}{*}{\cite{ospina2020feasibility}}  & Dyn. mode decomposition& Simulation-based &{Static} &Frequency instability \\ \cline{2-6}
    & \multirow{1}{*}{\centering\cite{dabrowski2017grid}} & {\centering $2^{nd}$ order model} & {\centering Simulation-based} & Static+Dyn. - Cont. load change &  {\centering Frequency instability}\\ \cline{2-6}
     & \multirow{1}{*}{\centering\cite{LaksRareEvent2022}}  & {\centering $3^{rd}$ order model} & {\centering Simulation-based} & {\centering Static} &Load/gen. shedding, Line disc. \\  \cline{2-6}
    &\multirow{1}{*}{\centering\cite{goodridge2024uncovering}}  & \multirow{1}{*}{$3^{rd}$ order model + $(N-1)$ constr.}& {\centering Simulation-based} & {Static+Dyn. - Discr. load change} & Load/gen. shedding, Line disc. \\ \cline{2-6}
     & \multirow{2}{*}{\centering\cite{Goodridgecascade2023}} & \multirow{2}{*}{$3^{rd}$ order model + $(N-1)$ constr.}& \multirow{2}{*}{\centering Simulation-based} &\multirow{2}{*}{\centering \makecell{ Static+Dyn. - Discr. load change}}  & Load/generation shedding, Line disc., Cascading failure \\ \cline{2-6}
      & \multirow{1}{*}{\centering\cite{acharya2020public}} & $1^{st}$ order model - DC PF approx. & {\centering Analytical} &{\centering Static} & {\centering Frequency instability}\\ \cline{2-6}
      & \multirow{1}{*}{\centering \cite{HammadSwitch2018}} & $1^{st}$ order model - DC PF approx. & {\centering Analytical} &{\centering Inter-area switching} & {\centering Frequency instability}\\ \cline{2-6}
      & \multirow{1}{*}{\centering \cite{AlanaziLoadOsc2023}} & $1^{st}$ order model - DC PF approx. & {\centering Analytical} & {\centering Inter-area switching} & {\centering Load shedding} \\ \hline
        
     \multirow{3}{*}{\rotatebox[origin = c]{90}{Dist.}}  &\multirow{1}{*}{\cite{LiuSoftOp2022}}  &  {Distflow} & Analytical & {Static} & Voltage constr. violation  \\ \cline{2-6}

    &\multirow{1}{*}{\cite{maleki2024impact}}  & {Lindistflow} &Analytical& {Static} & Voltage constr. violation  \\ \cline{2-6}

    &  \multirow{1}{*}{\cite{khan2019impact}} & {MATPOWER model} & Simulation-based &  {Static} & Voltage constr. violation \\ \hline
    {T\&D}& \multirow{1}{*}{\cite{dvorkin2017iot}} &Coordinated T\&D model & Simulation-based&{Static} & Socio-economic impacts\\ \hline

    \end{tabular}
\vspace{-2mm}
    \label{LAA_impacts}

\end{table*}

\vspace{-4mm}
\subsection{Impact on Energy Markets} 
\vspace{-1mm}

Grid stability can be severely impacted when attackers compromise a large number of loads, resulting in significant frequency deviations and voltage fluctuations that may lead to cascading failures or blackouts. However, the risks posed by LAAs extend beyond grid performance. In addition to disrupting grid operations, LAAs can also be leveraged for economic manipulation, which has emerged as a growing concern \cite{FMT, AUCPTS, Sophist}.  
The energy market, being a central component of financial transactions within the power system, is inherently vulnerable to exploitation by LAAs. By compromising and intentionally manipulating the demand and supply of critical loads, attackers can rapidly alter the supply-demand balance, triggering dramatic price swings and seeking economic profits. 

\subsubsection{Mechanism of Energy Market}
The energy market functions as a competitive platform where electricity generators offer their production to meet demand, while electric utilities and other market participants purchase electricity through real-time and day-ahead markets. In the day-ahead market, participants submit bids during the day-ahead bid period, followed by market clearing, where bids are processed, and energy schedules are set \cite{liu2015bidding}. In the intra-day market, participants refine their bids and adjust their positions in response to updated information. The real-time market focuses on real-time flexibility bids, allowing providers to adjust supply and demand \cite{wang2015review}. It balances the gaps between day-ahead commitments and the real-time electricity supply and demand in real time, which ensures grid stability by aligning supply with real-time demand. 

Electricity pricing is a fundamental aspect of energy markets. Common pricing schemes include flat pricing, time-of-use (ToU) pricing, and real-time pricing (RTP). While flat pricing sets a constant price throughout the day, ToU pricing categorizes the day into different time periods, each with a specific price that is predetermined and announced before the operation begins \cite{samadi2010optimal}. In contrast, RTP adjusts electricity costs based on actual market conditions in real-time, typically on an hourly or sub-hourly basis, reflecting fluctuations in supply and demand. This approach factors in grid conditions, resource availability, and consumption changes. By exposing consumers to real-time price variations, RTP promotes flexible consumption behaviors, which enhances load management and grid stability. It is especially beneficial with the integration of RES, as it aligns consumption with periods of high generation or lower demand, ensuring efficient grid use \cite{anand2018real}.

\subsubsection{LAAs to Disrupt the Energy Market Operation} 
Real-time electricity prices in energy markets are closely tied to system load demands \cite{roozbehani2012volatility}, making them susceptible to manipulation through LAAs. By compromising large-scale loads, attackers can influence prices or disrupt the market \cite{ospina2023feasibility, barreto2014cps,shekari2021mamiot,jahangir2024charge}. Understanding the impact of LAAs on the energy markets is crucial, as several studies demonstrate the feasibility of such attacks. Table \ref{LAA_market} summarizes existing research, which is further discussed below. 

\begin{table}
    \centering
    \vspace{-1mm}
    \caption{Literature of LAAs effect in energy market operations.}
    \vspace{-2mm}
    \resizebox{0.95\columnwidth}{!}{
    \begin{tabular}{|c|c|c|} \hline
    Target Market & ~Ref~ &  Target Devices\\   
    \hline
    \multirow{2}{*}{Real-time market} &\cite{ospina2023feasibility} &  MV/LV Loads     \\
    \cline{2-3}
    &\cite{barreto2014cps} &  Flexible Loads \\
    \hline
    Day-ahead \& Real-time  &\cite{shekari2021mamiot} &  ~High-Wattage IoT Botnets~ \\
    \cline{2-3}
    market&\cite{jahangir2024charge} &  EVCS \\
    \hline
    \end{tabular}}
    \vspace{-2mm}
    \label{LAA_market}
\end{table}

The manipulation of market via IoT (MaMIoT) attack, introduced in \cite{shekari2021mamiot}, leverages LAAs on high-wattage IoT botnets to affect the electricity prices in the energy market. An optimization model is proposed to solve for a stealthy and repeatable LAA, which maximizes the attack's impact. By increasing the power demand of the compromised devices, the RTPs can be intentionally increased, which in turn enables power generation utilities to gain more benefits. Results demonstrate that even with a small number of controllable bots, the attacker can significantly increase their profits while inflicting substantial financial losses on other market participants.

The authors in \cite{ospina2023feasibility} studied how LAAs affect locational marginal prices (LMPs) in distribution systems, showing that LAAs can propagate from targeted distribution systems through the transmission system to neighboring areas. Their findings reveal that LAAs at specific loads can substantially raise electricity prices locally. Moreover, energy storage system (ESS) owners can leverage LAAs to create high-loading conditions, enabling them to sell energy at inflated prices for profit. Similarly, Barreto \textit{et al.} \cite{barreto2014cps} examined demand response vulnerabilities, identifying two attacker types: malicious actors aiming to damage grid equipment and selfish attackers seeking profit. Results indicate that exploiting direct load control can significantly boost attacker profits.

The work in \cite{jahangir2024charge} investigated the charging manipulation attacks (CMAs) against EV charging, which aims to shift the EV aggregator's demand to different periods of the day. The study delved into the impact of CMAs on aggregators’ profit and performance in the day-ahead markets and real-time markets by modeling their participation in both markets. Results show that the CMAs can lead to a considerable increase in the aggregated surcharge charging cost. 

In conclusion, the energy market can be highly vulnerable to exploitation by LAAs, especially given the increasing reliance on wholesale electric energy pricing mechanisms like LMP. These attacks can manipulate electricity prices by artificially inflating or deflating demand, resulting in financial gains for attackers and potential market destabilization. It is essential to develop robust cybersecurity measures and prevent LAAs from manipulating energy prices.

\vspace{-2mm}
\section{Attack Detection and Localization} \label{detection}
\vspace{-1mm}
\begin{table}
    \centering
    \caption{Detection and localization approaches against LAAs.}
    \vspace{-2mm}\resizebox{0.95\columnwidth}{!}{
    \begin{tabular}{|c|c|c|} \hline
    Method Type & Ref & Approach \\   
    \hline
    \multirow{3}{*}{Model-Based}&\cite{mellucci2015load}&SMO + generator phase angle\\   
    \cline{2-3}
    &\cite{su2021observer,rinaldi2022load,e2022load,li2023dynamic} &SMO + frequency\&phase angle \\
    \cline{2-3} 
    &\cite{izbicki2017identification}&KF + frequency\&phase angle\\
    \hline
    
    \multirow{6}{*}{Data-Driven}&\cite{amini2015detecting}&W-FFT\&W-CC + load demand\\
    \cline{2-3}
    &\cite{youssef2022detection}&TDLL + load demand \\
    \cline{2-3}
    &\cite{sawas2023real}&LSTM + load demand \\
    \cline{2-3}
    &\cite{ebtia2024spatial}&GAT\&LSTM + load demand\\
    \cline{2-3}
    &\cite{jahangir2023deep}&CNN + frequency\&phase angle\\  
    \cline{2-3}
    &\cite{jahangir2024charge}&CNN + EV charging profiles \\
    \hline
    
    \multirow{4}{*}{Hybrid}&\cite{amini2017hierarchical}& MILP + frequency\&phase angle\\
    \cline{2-3}
    &\cite{amini2017hierarchical}& LQR + CNN + frequency\\
    \cline{2-3}
    &\cite{lakshminarayana2021datadriven}& SR + frequency\&phase\\
    \cline{2-3}
    &\cite{lakshminarayana2021datadriven}& PINN + frequency\&phase\\
    \hline
    \end{tabular}}
    \vspace{-6mm}
    \label{LAA_Detection}
\end{table}

If a successful LAA occurs, swift detection and localization are of utmost importance to limit the damage.
 Table \ref{LAA_Detection} summarizes the work in this area, classifying approaches into model-based, data-driven, and hybrid methods. Model-based approaches utilize predefined algorithms or mathematical models to detect and locate LAAs, relying on a thorough understanding of system dynamics. Data-driven approaches utilize ML algorithms to automatically capture potential relationships in the data and anomalies caused by LAAs. Hybrid approaches use a combination of the two.

\vspace{-4mm}
\subsection{Model-Based Approaches}
\vspace{-1mm}

One approach to detect LAAs is to model the power network as a semi-explicit class of differential-algebraic equations (DAEs) \cite{mellucci2015load} and then use a sliding mode observer (SMO) to estimate the state of the system. The observer is able to estimate any signals representing failures to detect LAAs, with only the generator phase angles being measured. Focusing on DLAAs, \cite{su2021observer} utilized the estimation of residual signals to detect attacks. Specifically, the power system was first modeled as state-space equations, and then a robust SMO was designed with an observer residual evaluation function. By comparing the value of this function to those observed in non-attack scenarios, the algorithm effectively identifies the DLAA.

Based on similar ideas of utilizing the observers, the sliding mode algorithm (STA) was used in the design of LAA detection while also incorporating the usage of ESS in the power system to defend against LAAs \cite{rinaldi2022load}. Another study utilized the linear matrix inequality (LMI) to design a bank of residual generators for functional observers to detect LAAs~\cite{e2022load}. This approach reduces the number of observed states, making it suitable for large-scale power network calculations. The researchers in \cite{li2023dynamic} designed a closed-loop SMO for the detection of closed-loop DLAAs, whose robustness is enhanced by using the $H_\infty$ performance index. A method using a low-rank Kalman filter was proposed in \cite{izbicki2017identification}, which identifies destabilizing attacks by estimating system states and attack parameters with the rank-1 method, achieving high accuracy and computational efficiency. 

Model-based approaches rely on observers or state estimators built using a state-space system model. While these methods provide strong theoretical foundations and require less data when the system is accurately modeled, the increasing integration of diverse equipment into power grids makes accurate modeling more challenging. This issue is particularly evident in large, complex grids, where developing precise state-space models requires significant effort, underscoring a key limitation of model-based approaches.

\vspace{-4mm}
\subsection{Data-Driven Approaches}
\vspace{-1mm}
With the rapid advancements in artificial intelligence-based techniques in recent years, several studies have adopted ML-based methods to detect and localize LAAs. As LAA affects the load profile of the system, an intuitive method is to monitor and analyze the load data recorded by smart meters. \cite{amini2015detecting} analyzed this data and showed that compromised loads can be detected by performing a fast Fourier Transform (FFT) and spectral analysis of the signals. A windowed FFT (W-FFT) and windowed cross-correlation (W-CC) were then used to detect an attack. They showed that utilizing time and frequency-domain signals could improve the performance of LAA detections. 

A time-delay neural network was designed in \cite{youssef2022detection}, whose inputs are the power demand profiles. Results showed that the detector could precisely recognize even LAAs that activate only a small number of electric water heaters. \cite{sawas2023real} pre-processed the load demand data into a feeder loading abnormal power spectrum index. This index was then fed into a long-short-term memory (LSTM) network to detect the anomalies. Another method was proposed in \cite{ebtia2024spatial} to detect LAAs in distribution networks by exploiting the spatial and temporal properties of the load data. A graph attention network (GAT) and LSTM were employed to capture the spatial and temporal correlations, respectively. The proposed model outperformed existing methods in detecting and localizing LAAs, demonstrating robustness against noise and outliers. 

It may not always be feasible to measure every real-time load profile. Existing work suggests that the sampling rate of meters could be exploited to launch stealthy load attacks. For instance, if attackers know the meter's sampling rate, they can synchronize the switching of power circuits to occur between meter readings, leading to inaccurate power consumption calculations \cite{wu2018false}. An alternative is to leverage the grid frequency and phase angle fluctuations monitored by PMUs installed to detect and localize LAAs. Based on this, \cite{jahangir2023deep} proposed the use of a convolutional neural network (CNN) classifier and reconstruction decoder to detect and localize DLAAs. The proposed method demonstrated good performance, even in the presence of noise or data loss in the measurements. Focusing on the LAAs utilizing EVCS, an unsupervised deep learning-based mechanism using CNN and EV charging profiles was proposed in \cite{jahangir2024charge} to detect malicious attacks.

Data-driven approaches reduce the reliance on accurate system modeling but come with their own challenges. Their effectiveness depends on the quality and representativeness of the training data, which may not fully capture all potential LAAs in real-world scenarios, limiting their feasibility. Additionally, neural networks, commonly used in these methods, face issues with interpretability, making it difficult to ensure robustness and security. Furthermore, data-driven algorithms are vulnerable to adversarial ML techniques \cite{youssef2023adversarial, sayghe2020survey, sayghe2020evasion, tian2022adversarial}, where malicious inputs can manipulate model predictions, further undermining their reliability.

\vspace{-4mm}
\subsection{Hybrid Approaches}
\vspace{-1mm}

\begin{figure}
    \centering    \includegraphics[width=0.99\linewidth]{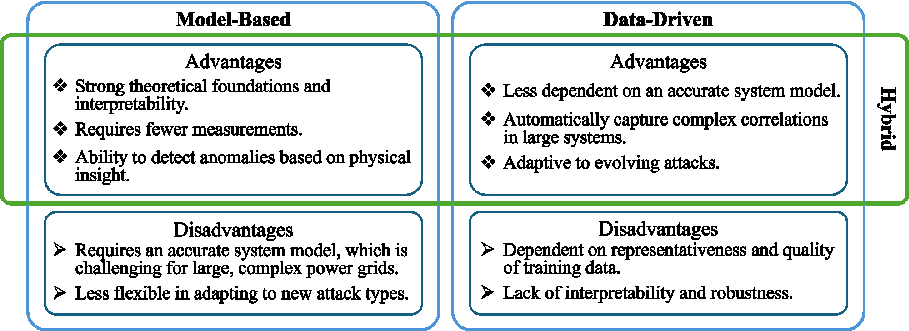}
    \vspace{-1mm}
    \caption{{Comparison among model-based, data-driven and hybrid approaches.}}
    \vspace{-4 mm}
    \label{fig:Hybrid}
\end{figure}

Given the shortcomings of both data-driven and model-based methods, recent work has explored hybrid approaches that combine the strengths of both, as illustrated in Fig. \ref{fig:Hybrid}. These approaches integrate the automatic fitting capabilities of data-driven methods with the physical insights provided by model-based techniques, offering more balanced and effective solutions.

Inspired by the model-based approaches,  \cite{amini2017hierarchical} followed the idea of designing observers to estimate the state of the system. The power system dynamics under attacks were represented using a state-space model, which was then transformed into the frequency domain. An optimization problem constrained by the frequency domain system model was then formulated in order to localize the attacks. The authors also pointed out the feasibility of applying the approach in wide area monitoring systems (WAMS), by implementing it in a hierarchical fashion.

A hybrid physics-based deep learning mechanism for LAA detection and localization was developed in \cite{sayed2024hybrid}. An observer based on a linear quadratic regulator (LQR) was designed to recover the states of the power system from generator frequencies, with which the pattern of load variations can be reconstructed by reversing the state-space equations. These reconstructed load patterns were then fed into a designed CNN classifier to identify whether and where the LAAs are.

Based on physics‐informed ML algorithms,  \cite{lakshminarayana2021datadriven} proposed two data-driven methods for LAA detection and localization. With phase angles and frequencies as the inputs, the authors designed algorithms based on sparse identification of non-linear dynamics and a physics‐informed neural network (PINN). Both algorithms incorporated the physical characteristics of the power system into the design, enabling the algorithm to operate without relying on offline training. 

\vspace{-4mm}
\section{Attack Mitigation}\label{Mitigation}
\vspace{-1mm}
Once detected, the destabilizing effects of LAAs need be addressed.
Attack mitigation schemes can be either preventive or reactive. Preventive methods aim at deploying security reinforcements \emph{offline} such that the attackers cannot launch destabilizing LAAs. For instance, reinforcing security features of load devices (e.g., adding hardware or software security features) can prevent load compromises and thus limit the attacker's ability to launch destabilizing LAAs. However, deploying these security measures may incur significant economic costs upfront. Thus, the second category of reactive methods aims to develop \emph{online} mitigating measures that will enhance the ability of power grids to withstand the adverse impacts of LAAs. This could, for instance, include generator dispatch from fast-acting resources to correct the imbalance caused by LAAs. Third, hybrid approaches include a combination of \emph{offline} deployments and \emph{online} operation of these devices to counter the adverse effects of LAAs. Table \ref{mitigation_table} summarizes the existing mitigation methods.  

\begin{table}
    \centering
    \caption{Classification of mitigation methods.}
    \vspace{-2mm}
    \begin{tabular}{|c|c|c|>{\centering\arraybackslash}p{3.6cm}|} \hline
    Type& Grid& Ref & Defensive tool \\   
    \hline 
  \multirow{11}{*}{\rotatebox[origin = c]{90}{Preventive}} & \multirow{8}{*}{Transmission} &\cite{amini2016dynamic}& Securing loads\\ \cline{3-4}
   &&\multirow{4}{*}{\cite{shekari2021mamiot}~} & Increasing data privacy, Sharing altered data, and installing non-intrusive load monitoring algorithm\\ \cline{3-4}
   & &\cite{soltan2019protecting} & Robust operating points\\ \cline{3-4}
   & &\cite{di2017optimization} &Battery\\ \cline{3-4}
   & &\cite{an2023robust}& Reactive power compensation\\  
   
   \cline{2-4}
   & \multirow{3}{*}{Distribution}&\cite{Shrestha_DataCentric2020} &Edge computing infrastructure \\\cline{3-4}
   && \cite{bellizio2022transition} & ML-based market clearing\\ \cline{3-4}
   &&  \cite{khan2021cyber}& Securing critical aggregators\\

    \hline
    
    \multirow{7}{*}{\rotatebox[origin = c]{90}{Reactive}} & \multirow{5}{*}{Transmission} &\cite{sayed2023protecting}  &EV charge/discharge\\ \cline{3-4}
   & & \cite{ChuMitigationTSG2023}&Frequency droop control \\ \cline{3-4}
   & & \cite{guo2021reinforcement}  &Load shedding\\  \cline{3-4}
    && \cite{gotte2022ripples} &Communication links \\ \cline{3-4}
    && \cite{barreto2020attacking} & Canceling selected bids\\
    \cline{2-4}
   &\multirow{2}{*}{Distribution}& \cite{maleki2024distribution} &Reconfiguration\\ \cline{3-4}
   && \cite{andriopoulos2024cyber} & Fraud data detection\\

        \hline 
        
   \multirow{4}{*}{\rotatebox[origin = c]{90}{Hybrid}}& \multirow{3}{*}{Transmission} & \cite{chen2020load} & Secondary frequency control\\ \cline{3-4}
    &&\multirow{2}{*}{\centering\cite{zhao2024integrated}}~ &Securing devices \& Optimal load management \\  \cline{2-4}
   &Distribution& \cite{LiuSoftOp2022} &Soft open points \\

    \hline 
    \end{tabular}   \label{mitigation_table}
\vspace{-6mm}

\end{table}

\vspace{-4mm}
\subsection{Preventive Mitigation}
\vspace{-1mm}

To minimize the protection cost, the amount and the location of the protected load must be selected strategically. \cite{amini2016dynamic} proposed a non-convex pole placement optimization problem to determine the optimal load protection to ensure that the attacker cannot launch destabilizing DLAAs even if the remaining vulnerable loads are compromised. 
The formulation ensures that the poles of the system (modeling the power grid frequency dynamics) always remain on the left half of the complex plane. The optimization is solved using a coordinated descent method. The work in \cite{LakshIoT2021} reformulated the load protection problem by modeling the system stability constraints using the eigenvalue sensitivity approach, which provides a linear approximation to the system's eigenvalues as a function of the attack parameters. The reformulated load protection problem becomes a simple linear programming problem, that significantly reduces the computational complexity. 

Different from load protection, 
\cite{soltan2019protecting} proposed an alternative approach to ensure system stability under DLAAs by finding \textit{robust generator operating points} using two approaches. The first approach aims to find generator operating points that incur the minimum cost such that no lines are overloaded following the generator's primary frequency control response (to an LAA). In cases where this condition cannot be ensured (due to insufficient generator capacities), they propose another approach which allows temporary line overflows that can eventually be cleared using the generator's secondary control. Furthermore, they provide upper and lower bounds on the magnitude of LAA that the grid is resilient against using the proposed primary and secondary robust operating points. Distributed storage units can also act as backup units for quick dispatch or absorption of the additional power demand. \cite{di2017optimization} proposed the optimal sizing and location of battery storage units accounting for security constraints under DLAA using the Lyapunov stability theorem. The problem is formulated as a nonconvex optimization problem and solved using a two-step solution inspired by the coordinate decent method. 

Game-theoretic approaches have been explored to model the strategic interactions between attackers and defenders in power grids. These interactions can be either preventive or reactive, depending on the operator's defensive strategy. For instance, when the defender anticipates a strategic attacker, preventive defenses can be designed accordingly. The work in \cite{an2023robust} presents a zero-sum Stackelberg game, where the defender acts as the leader and the attacker as the follower. In this formulation, the attacker manipulates the reactive power set points at load buses to compromise voltage stability, while the defender installs reactive power compensation (RPC) devices to minimize the attack's impact.

To reduce the risk of successful LAAs, \cite{Shrestha_DataCentric2020} proposed an edge-computing infrastructure that dynamically enforces permissions for each IoT device (e.g., read and execute commands), preventing unauthorized control commands on critical devices and minimizing disruption. Edge servers near IoT devices reduce latency and enable quick responses to attacks.

While most existing studies focus on mitigating the destabilizing effects of LAAs, a few works focus on mitigating LAA impacts on the electricity market or market-based mitigation. \cite{bellizio2022transition} uses an LSTM-based method to capture temporal patterns the energy prices based on the import and export prices set by retailers and communicated to households. Once the LTSM is trained, a confidence interval is established based on prediction errors, and any price prediction that falls outside this range is flagged as a potential attack. \cite{shekari2021mamiot} provides information on the most effective countermeasures against market manipulation attacks, including sharing detailed market data only with market players such as independent system operators, releasing redacted or altered versions of market data or even delaying the release of full data sets to the public. These methods limit access to historical market data and enhance data privacy. As a result, implementing real-time attacks against the market becomes difficult. Furthermore, installing non-intrusive load monitoring (NILM) at the household level can be an effective countermeasure. For example, NILM can detect the unusual operation of home appliances during the periods when the homeowner is not present. \cite{khan2021cyber} proposes a mixed-integer nonlinear 
programming optimization problem to design an attack against the electricity market that causes the highest tariff increment and congestion issues in the network. Based on this formulation, they locate the aggregators from which the attack can be launched with the least effort and propose securing these aggregators to mitigate the attacks.

\vspace{-4mm}
\subsection{Reactive Mitigation}
\vspace{-1mm}
Reactive attack mitigation refers to restorative actions following the detection of an LAA. The mitigation scheme proposed by \cite{ChuMitigationTSG2023} leverages the high integration of inverter-based resources (IBRs) to mitigate LAA. The main idea is to dispatch power from fast-acting IBRs to counter the imbalance caused by LAAs following attack detection (e.g., using the ML approaches listed in Section~\ref{detection}). This is achieved by optimizing the droop control parameters of IBRs within an economic dispatch framework. To account for the uncertainty associated with attack detection results, the authors develop a distributionally robust optimization approach.

The game-theoretic approach has also been used for reactive defense where the operator acts as a follower to mitigate the attack. In \cite{guo2021reinforcement}, the authors developed a two-player zero-sum multistage game-theoretic formulation. The attacker's action corresponds to load changes at different time instants of a DLAA, whereas the defender's action corresponds to load shedding to prevent the attack's destabilizing effects. To find the optimal sequence of actions, a minimax-q learning-based solution is proposed. However, the solution requires the knowledge of the action space of players, objective functions, and the system's dynamics.

When conventional communication media such as the internet or cellular networks are compromised during a cyber attack, having a backup communication link can provide essential control pathways to send mitigation commands, such as disconnecting compromised loads. \cite{gotte2022ripples} explored using the power grid frequency as an alternative communication medium by modulating the power consumption of large grid-connected loads or generators. For instance, a large consumer like an aluminum smelter can be modified to act as an on-demand grid frequency modulation transmitter at a small cost.
This creates small, controlled changes to the grid's frequency which can be used to transmit load disconnection messages.

The flexibility provided by EVs is proposed to mitigate LAAs in \cite{sayed2023protecting}. The EV control actions (such as charge/discharge actions) to compensate for the LAA imbalance are optimized using $H-2$ and $ H-\infty$-based robust control techniques. The work also takes the uncertainties like EV users' behavior into account. The proposed scheme keeps the system's frequency in the stable range, thus preventing subsequent contingencies such as line tripping or generator relay tripping. 

A Stackelberg game approach is proposed in \cite{maleki2024distribution} to mitigate LAAs in distribution systems, with the attacker as the leader executing LAA and network reconfiguration as a defensive action. As noted in Section~\ref{Impact_stability}, in a distribution network, the location of the LAA plays a key role in determining the attack impact \cite{maleki2024impact}. By network reconfiguration, the defender can potentially alter the location of the LAA within the distribution network, thus reducing the detrimental effect of the attack. 

To counteract the attacks on electricity markets, \cite{andriopoulos2024cyber} propose a scheme to detect and penalize agents that manipulate the electricity market by introducing a fraud detection tool that compares real-time bids with historical data. Any agent (node) identified as an adversary will incur a penalty. Furthermore, \cite{barreto2020attacking} considers that the operator has historical data of the bids submitted by the market players. Based on this when there is an attack, they can estimate the number of devices under attack as well as the impact of the attack. To compensate for the impact of the attack on electricity price (which is the increment in the price), the operator drops some of the highest-priced bids to correct the effect of the attack.

\vspace{-5mm}
\subsection{Hybrid Mitigation}
\vspace{-1mm}
Finally, we discuss hybrid mitigation strategies that involve jointly optimizing offline deployments and online operations. \cite{chen2020load} introduced a secondary frequency control approach that operates both preventively and reactively. The preventive method uses a model-free approach based on RL in which the system learns various attack strategies offline and saves the optimal defense strategies. When an attack is detected, the algorithm matches the current scenario with the pre-trained strategies to find the best mitigation action. This approach actively adjusts to specific attacks and uses feedback control to minimize system disturbances.

In \cite{zhao2024integrated}, the work explored a game-theoretic defense that integrates both the cyber and physical aspects of the power grid. {At the cyber layer, the attacker seeks to spread malware in IoT devices, while the defender focuses on securing them (e.g., through software and firmware updates).} At the physical layer, the attacker aims to launch LAAs that maximize the impact of the attack, while the defender counteracts it by optimal load management and control strategies to maintain the stability of the grid. Results show that the formulation enables the operator to maintain the system's normal operation under the strategic botnet attack and improve the grid's integrative resiliency at both cyber and physical layers. 

\cite{LiuSoftOp2022} proposed the use of SOPs in distribution networks that provide the flexibility of connecting different buses when required and controlling the active/reactive power flows \cite{jiang2022overview}. 
The SOP installation is performed offline and involves solving a chance constraint optimization problem considering all possible LAA scenarios (i.e., the attack location and the magnitude of the compromised load). 
The SOP operation is computed in real-time following an attack by solving a second-order conic programming optimization. The objective is to minimize the grid's nodal voltage deviation from the nominal value under the power flow and SOPs operation constraints.

\vspace{-3mm}
\section{Conclusions \& Future Research} \label{conclusion_future_research}

\vspace{-2 mm}
LAAs pose a significant cybersecurity threat to modern power grids, exploiting IoT-enabled high-wattage consumer devices to manipulate power consumption at scale. This paper provides a comprehensive introduction to LAAs by classifying them based on various evaluation criteria, offering a deeper understanding of their forms and impact. The survey also systematically investigates the impacts of LAAs on power system operations, revealing the primary consequences: grid frequency and voltage instability, and electricity market disequilibrium. A comparative analysis of detection methodologies is conducted, encompassing model-based analytical techniques, data-driven machine-learning approaches, and hybrid detection frameworks. The mitigation framework is comprehensively evaluated through two complementary dimensions: 1) preventive security measures, including cryptographic protection mechanisms for IoT-enabled smart loads, and 2) reactive countermeasures incorporating adaptive demand-side management protocols and real-time emergency load shedding algorithms.

The evolving landscape of LAAs presents significant cybersecurity risks to power systems, necessitating proactive policy interventions, regulatory frameworks, and industry practices. The revealed vulnerability of IoT-enabled high-wattage devices underscores the urgency for implementation of robust security measures, such as IEEE 2030.5 (Smart Energy Profile 2.0) \cite{passos2025tutorial} security extensions, which provides secure communication mechanisms for DERs. Moreover, recent governmental responses highlight the importance of securing smart appliances and demand-side response mechanisms to prevent large-scale cyber threats. For instance, the Cyber Resilience Act by the European Commission aims to enhance the security of digital products by enforcing stricter cybersecurity requirements across the supply chain \cite{cyberResilienceAct}. Similarly, in the U.S., the NIST cybersecurity labeling program \cite{NISTLabeling} focuses on improving consumer awareness and incentivizing manufacturers to implement stronger security measures in consumer devices, contributing to the overall resilience of energy systems. UK’s smart energy security framework also emphasizes securing demand-side response technologies and ensuring compliance with cybersecurity standards for energy smart appliances \cite{UKDelivering}.

To effectively mitigate LAA risks, it is imperative that policymakers establish enforceable cybersecurity baselines that integrate LAA-specific threat models into existing grid security frameworks. Ensuring compliance with standardized security protocols, such as IEEE 2030.5, plays a crucial role in enhancing interoperability and protecting IoT-enabled loads from large-scale coordinated LAAs. Additionally, industry stakeholders should prioritize the deployment of advanced anomaly detection mechanisms and adaptive defense strategies to counter evolving threats. A multidisciplinary approach that combines regulatory enforcement, technological innovation, and cross-sector collaboration is essential to enhancing grid resilience against cyber-physical threats. By integrating these measures, power systems can transition toward a more secure, intelligent, and adaptive operational paradigm, effectively mitigating the impact of LAAs while maintaining system stability and reliability.

Several open research problems remain, offering fertile ground for future exploration. 

{\it(1) Joint Cyber-Physical Risk Analysis:}
Although several works have analyzed the risks associated with LAAs, there remains a gap in understanding the joint cyber-physical impacts, particularly in the context of DLAAs. Existing research has primarily focused on either cyber or physical aspects, but an integrated analysis is crucial. For example, \cite{zhao2024integrated} proposed a two-stage framework where they first analyzed cyber-risks using an epidemic model to track botnet propagation, followed by an examination of the physical impacts on the power grid. Future work should assess cyber-physical risks and {develop optimized} joint defense strategies, particularly in systems with high IoT and DER penetration.

{\it (2) LAA Market Manipulation and Economic Impacts:}
LAAs present significant threats to energy markets by enabling attackers to manipulate market prices through changes in power demand. These attacks can lead to extreme volatility, with prices surging or plummeting based on the attacker’s intent. While the feasibility of manipulating energy markets through LAAs has been demonstrated \cite{ospina2023feasibility}, the economic impacts, real-time market response mechanisms, and security protocols to ensure market integrity during and after LAAs, remain underexplored areas.

{\it (3) LAAs in Renewable Energy-Rich Systems:}
The inherent variability and intermittency of RES make the power grid more susceptible to LAAs. For instance, an attacker could exploit periods of low RES generation by launching an LAA \cite{LakshCOVID2022}, triggering an increase in demand that the grid cannot meet due to insufficient backup from conventional generators. Similarly, during periods of high RES output, LAAs could result in overproduction, leading to instability in grid frequency and voltage. Future research should focus on integrating advanced forecasting techniques and real-time monitoring which could help grid operators maintain stability during LAAs.

{\it (4) Cyber-Resilient Recovery Strategies:}
As demonstrated in \cite{liu2023cyber}, effective recovery from cyber-physical attacks like DLAAs requires the development of integrated cyber-physical recovery frameworks. The proposed cyber recovery from DLAAs focuses on coordinating repair crews and dynamically adjusting IBR droop gains to restore stability. Future work should build upon this by exploring how similar recovery frameworks can be tailored for LAA scenarios, particularly in RES-dominated grids. Research should also investigate real-time adaptive recovery strategies that leverage ML to predict attack evolution and optimize repair and stabilization efforts.

{\it (5) Enhanced Detection and Mitigation via Game Theory and ML:}
Given the evolving sophistication of LAA strategies, future research should focus on developing more robust detection and mitigation mechanisms. Game-theoretic approaches, as in \cite{liu2023cyber, maleki2024distribution, ghosh2024bi}, can model the interaction between attackers and defenders, enabling the prediction of optimal attacks and defenses. Additionally, combining game theory with ML to identify abnormal power demand patterns with physics-based models to maintain system stability presents a promising approach for enhancing LAA defense strategies.

\vspace{-2mm}
\section*{Acknowledgments}
\vspace{-1mm}
This work has been supported in part by PhD Cofund WALL-EE project between the University of Warwick, UK and CY Cergy Paris University, France, EPSRC under Grant EP/S035362/1, and King Abdullah University of Science and Technology (KAUST) under Awards No. ORFS-2022-CRG11-5021 and No. RFS-OFP2023-5505.

\vspace{-3mm}
\balance
\bibliographystyle{IEEEtran}
\bibliography{IEEEabrv,bibliography}

\end{document}